\documentclass[twocolumn,showpacs,preprintnumbers,amsmath,amssymb,prl,superscriptaddress]{revtex4-1}
\usepackage{graphicx}
\usepackage{dcolumn}
\usepackage{mathrsfs}
\usepackage{bm}
\usepackage{color}

\begin{document}

\preprint{}

\title{Bell inequality of frequency-bin entangled photon pairs with time-resolved detection}

\author{Xianxin Guo}
\affiliation{Department of Physics, The Hong Kong University of Science and Technology, Clear Water Bay, Kowloon, Hong Kong, China}

\author{Yefeng Mei}
\affiliation{Department of Physics, The Hong Kong University of Science and Technology, Clear Water Bay, Kowloon, Hong Kong, China}

\author{Shengwang Du}\email{Corresponding author: dusw@ust.hk}
\affiliation{Department of Physics, The Hong Kong University of Science and Technology, Clear Water Bay, Kowloon, Hong Kong, China}

\date{\today}

\begin{abstract}
{Violation of Bell inequality, for ruling out the possibility of local hidden variable theories, is commonly used as a strong witness for quantum entanglement. In previous Bell test experiments with photonic entanglement based on two-photon coincidence measurement, the photon temporal wave packets are absorbed completely by the detectors. That is, the photon coherence time is much shorter than the detection time window. Here we demonstrate generation of frequency-bin entangled narrowband biphotons, and for the first time, test the Clauser-Horne-Shimony-Holt (CHSH) Bell inequality $|S|\leq 2$ for their nonlocal temporal correlations with time-resolved detection. We obtain a maximum $|S|$ value of $2.52\pm0.48$ that violates the CHSH inequality. Our result will have applications in quantum information processing involving time-frequency entanglement.}
\end{abstract}





\maketitle


\textit{Introduction.---}As one of the most important features of quantum mechanics, entanglement is essential in quantum information processing, quantum computation, and quantum communication \cite{Entanglement}. Photonic entanglement has been realized in diverse degrees of freedom, including polarization \cite{OuPRL1988,ShihPRL1988,KwiatPRL1995}, position-momentum \cite{HornePRL1989,RarityPRL1990}, orbital angular momentum \cite{MairNature2001}, and time-frequency \cite{FransonPRL1989, StefanovPRA2013, OlislagerPRA2010, XieNature2015, RamelowPRL2009, GisinNP2007}. The nonlocal correlations between distant entangled photons provide a standard platform for Bell test \cite{Bell1964, BellBook, CHSH1969}, and confirm that quantum mechanics is incompatible with the local hidden variable theories which is the essence of the Einstein-Podolsky-Rosen (EPR) paradox \cite{EPR}. Recently loophole-free test of Bell's theorem has been demonstrated \cite{LoopholeFreeTest}. On the other side, violation of Bell inequality is often used as an entanglement witness.

In most previous experimental tests of Bell's theorem with entangled photons, the photon wave packets are absorbed completely by the coincidence detectors. That is, the photon coherence time is much shorter than the detection time window. In these works, the photon coincidence detection is modelled as integral over the entire wave packets. Surprisingly, although the time-frequency entanglement has been intensively studied \cite{FransonPRL1989, StefanovPRA2013, OlislagerPRA2010, XieNature2015, RamelowPRL2009, GisinNP2007}, the nonlocal correlation between the arrival times on the detectors of frequency-bin entangled photons (or the biphoton temporal correlation) has never been used for testing Bell's theorem.

Recent development of narrowband biphoton generation makes it possible to reveal the rich temporal quantum state information directly with time-resolved single-photon counters. Producing biphotons with a bandwidth narrower than 50 MHz has been demonstrated with spontaneous parametric down conversion insider a cavity \cite{FeketePRL2013, PanPRL2008}, spontaneous four-wave mixing (SFWM) in a hot atomic vapor cell \cite{Du2016} or laser-cooled atoms \cite{DuPRL2008, ZhaoOptica2014, ChenSR2015, KurtsieferPRL2013, YanPRL2014, KimPRL2014}. Although the time-frequency entanglement is naturally endowed by the energy conservation in generating these narrowband biphotons \cite{TQST2015}, its violation of Bell inequality has not been directly tested.

In this Letter, we demonstrate generation of frequency-bin entangled narrowband biphotons using SFWM in cold atoms for testing Bell's theorem. For the first time, we test the Clauser-Horne-Shimony-Holt (CHSH) Bell inequality $|S|\leq 2$ for their nonlocal temporal correlations with time-resolved detection. We obtain a maximum $|S|$ value of $2.52\pm0.48$ that violates the CHSH inequality. Our result also reveals the connection between the visibility of the two-photon quantum temporal beating resulting from the frequency entanglement and the violation of the Bell inequality.

\begin{figure*}
\includegraphics[width=15cm]{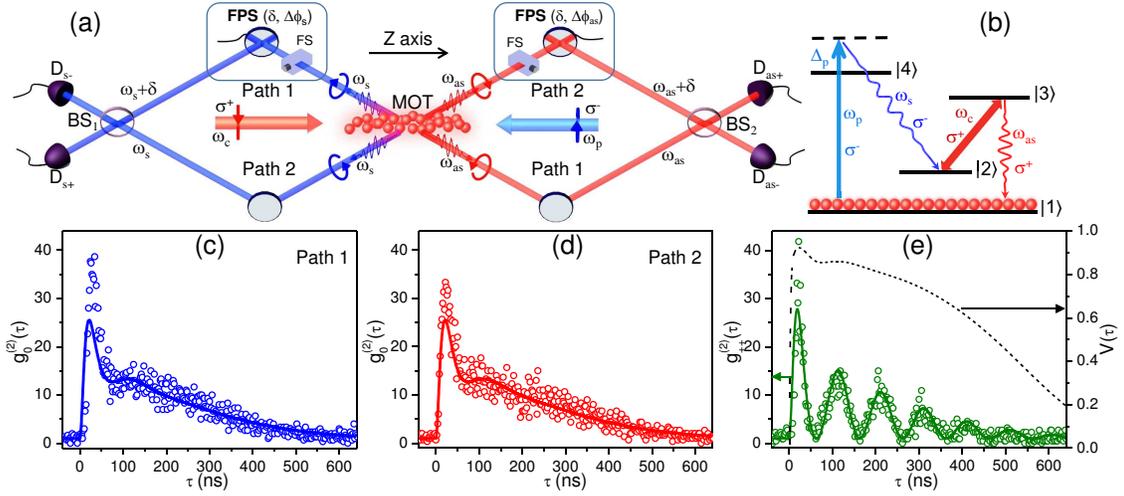}
\caption{\label{fig:schematic} (color online). Generation of frequency-bin entangled narrowband biphotons. (a) Experimental schematics of generating biphotons with double-path spontaneous four-wave mixing (SFWM) from cold $^{85}$Rb atoms. Backward and paired Stokes ($\omega_\mathrm{s}$) and anti-Stokes ($\omega_\mathrm{as}$) photons are spontaneously produced into paths 1 and 2, which are symmetric with angles of $\pm3^{\circ}$ to the longitudinal axis. The Stokes photons in path 1 go through a frequency-phase shifter FPS($\delta$, $\Delta\phi_s$), and the anti-Stokes photons in path 2 go through FPS($\delta$, $\Delta\phi_\mathrm{as}$).  BS$_1$ and BS$_2$ are two beam splitters. D$_\mathrm{s\pm}$ and D$_\mathrm{as\pm}$ are single-photon counting modules. (b) $^{85}$Rb atomic energy level diagram for SFWM. The atomic hyperfine levels are chosen as $|1\rangle=|5S_{1/2}, F=2\rangle$, $|2\rangle=|5S_{1/2}, F=3\rangle$, $|3\rangle=|5P_{1/2}, F=3\rangle$, and $|4\rangle=|5P_{3/2}, F=3\rangle$. The circularly polarized ($\sigma^{-}$) pump laser (780 nm) is blue detuned by 60 MHz from the transition $|1\rangle\rightarrow|4\rangle$, and the coupling laser ($\sigma^{+}$, 795 nm) is on resonance to the transition $|2\rangle\leftrightarrow|3\rangle$. (c) and (d) are the biphoton temporal correlations in paths 1 and 2, respectively. (e) Biphoton quantum beating measured between the detectors D$_\mathrm{s+}$ and D$_\mathrm{as+}$ and its visibility with the phase setting $\Delta\phi_\mathrm{s} =3\pi/2$ and  $\Delta\phi_\mathrm{as} = -\pi/4$. The solid curves in (c), (d), and (e) are obtained from the SFWM biphoton theory \cite{SupplementalMaterial}.}
\end{figure*}

\textit{Generation of frequency-bin entangled narrowband biphotons.---} Our experimental configuration of double-path SFWM and relevant atomic energy level diagram are illustrated in Figs. \ref{fig:schematic}(a) and \ref{fig:schematic}(b). We work with laser-cooled $^{85}$Rb atoms trapped in a two-dimensional (2D) magneto-optical trap (MOT) \cite{2DMOT}. In the presence of counter-propagating pump ($\omega_\mathrm{p}$) and coupling ($\omega_\mathrm{c}$) laser beams along the longitudinal z axis of the 2D MOT, correlated Stokes ($\omega_s$) and anti-Stokes ($\omega_{as}$) photons are spontaneously generated in opposite directions and collected into two spatial symmetric single-mode paths (paths 1 and 2). We send the Stokes photons in path 1 through a frequency-phase shifter FPS($\delta=2\pi \times 10$ MHz, $\Delta\phi_\mathrm{s}$) so that these Stokes photons' frequency become $\omega_\mathrm{s}+\delta$ and obtain a relative phase $\Delta\phi_\mathrm{s}$ to those in path 2. Symmetrically, we add a second FPS($\delta, \Delta\phi_\mathrm{as}$) to the anti-Stokes photons in path 2.  We then combine the two paths with two beam splitters (BS$_1$ and BS$_2$, 50\% : 50\%). The single-mode outputs of the beam splitters are detected by single-photon counting modules (D$_\mathrm{s\pm}$ and D$_\mathrm{as\pm}$), as shown in Fig. \ref{fig:schematic}(a). The detailed description of our experimental set-up is presented in the Supplemental Material \cite{SupplementalMaterial}. The photon pairs from the beam-splitter outputs are frequency-bin entangled and their biphoton states are described as
\begin{eqnarray}
|\Psi_\mathrm{XY}\rangle &=& \frac{1}{\sqrt{2}}\big[|\omega_\mathrm{s}+\delta\rangle|\omega_\mathrm{as}\rangle \nonumber\\
&+& \mathrm{XY} e^{i(\Delta\phi_\mathrm{as}-\Delta\phi_\mathrm{s})}|\omega_\mathrm{s}\rangle|\omega_\mathrm{as}+\delta\rangle\big], \label{eq:Entangled state}
\end{eqnarray}
where $\mathrm{XY}$, as a product of the signs (+, -), represents the Stoke to anti-Stokes combinations from the beam splitter outputs: $|\Psi_\mathrm{XY}\rangle$ is detected by (D$_\mathrm{sX}$, D$_\mathrm{asY}$).

\begin{figure*}
\includegraphics[width=17cm]{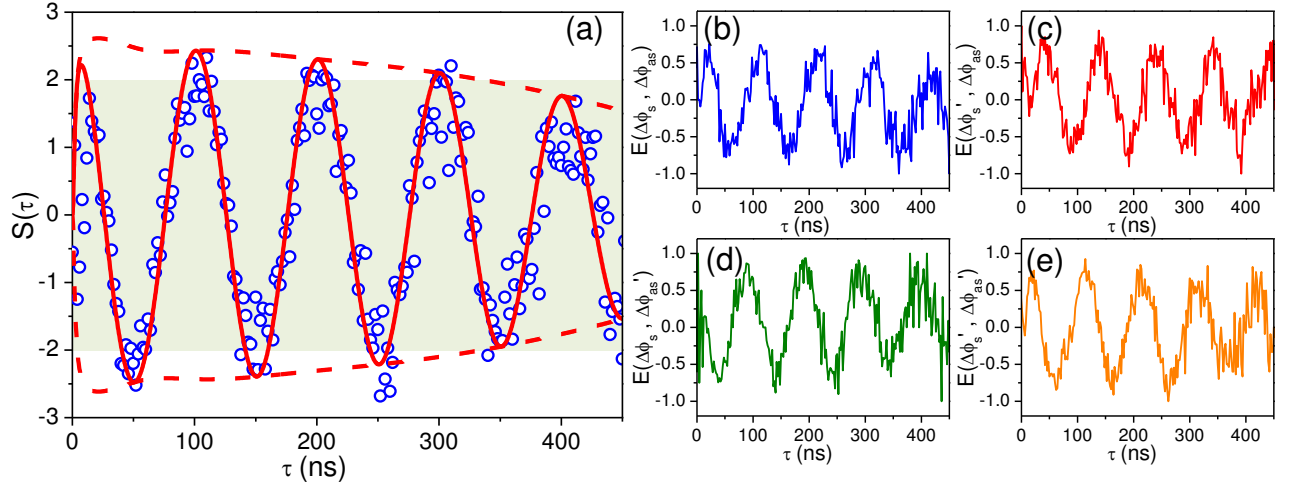}
\caption{\label{fig:Bell inequality} (color online). CHSH Bell inequality of frequency-bin entanglement. (a) $S$ as a function of two-photon relative time delay $\tau$. The circles are experimental data. The solid curve is predicted by the theory. The dashed envelopes are plotted from $\pm2\sqrt{2}V(\tau)$, where the visibility $V(\tau)$ is determined by Eq. (\ref{eq:V}). The shadow area is the classical regime where $|S|\leq 2$. (b)-(e) are the measured Bell correlations. The phase settings are $\Delta\phi_\mathrm{s} = 0$, $\Delta\phi_\mathrm{as} =\pi/4$, $\Delta\phi_\mathrm{s}' = -\pi/2$, and $\Delta\phi_\mathrm{as}' = -\pi/4$.}
\end{figure*}

The Glauber correlation function of the biphoton state in  Eq. (\ref{eq:Entangled state}) exhibits a quantum beating \cite{SupplementalMaterial}:
\begin{eqnarray}
G_\mathrm{XY}^{(2)}(&\tau&; \Delta\phi_\mathrm{s}, \Delta\phi_\mathrm{as})= \frac{1}{2} [G^{(2)}_0(\tau)-N_0]\nonumber\\
&& \times [1+\mathrm{XY} \cos(\delta\tau+\Delta\phi_\mathrm{s}-\Delta\phi_\mathrm{as})]+N_0,
\label{eq:G2 function}
\end{eqnarray}
where $\tau=t_\mathrm{as}-t_\mathrm{s}$. $G^{(2)}_0(\tau)$ is the biphoton Glauber correlation function before the beam splitters, which is the same for both paths 1 and 2. $N_0$ is the uncorrelated accidental coincidence rate. We then have the normalized biphoton correlation function
\begin{eqnarray}
&&g_\mathrm{XY}^{(2)}(\tau; \Delta\phi_\mathrm{s}, \Delta\phi_\mathrm{as})=G_\mathrm{XY}^{(2)}(\tau; \Delta\phi_\mathrm{s}, \Delta\phi_\mathrm{as})/N_0 \nonumber\\
&&=\frac{1}{2} [g^{(2)}_0(\tau)-1][1+\mathrm{XY} \cos(\delta\tau+\Delta\phi_\mathrm{s}-\Delta\phi_\mathrm{as})]+1, 
\label{eq:g2 function}
\end{eqnarray}
where $g^{(2)}_0(\tau)=G^{(2)}_0(\tau)/N_0$. As $N_0>0$, the beating visibility slowly varies as a function of $\tau$:
\begin{eqnarray}
V(\tau)=\frac{G^{(2)}_0(\tau)-N_0}{G^{(2)}_0(\tau)+N_0}=\frac{g^{(2)}_0(\tau)-1}{g^{(2)}_0(\tau)+1}.
\label{eq:V}
\end{eqnarray}
If $N_0=0$, the cosine modulation in the quantum beat has a visibility of 100\%. Experimentally, the two-photon temporal correlation is measured as coincidence counts between the detectors
\begin{eqnarray}
C_\mathrm{XY}(\tau; \Delta\phi_\mathrm{s}, \Delta\phi_\mathrm{as}) = G_\mathrm{XY}^{(2)}(\tau; \Delta\phi_\mathrm{s}, \Delta\phi_\mathrm{as})\eta\xi\Delta t_\mathrm{bin}T,
\label{eq:Coincidence}
\end{eqnarray}
where $\eta$ is the joint detection efficiency, $\xi$ is the duty cycle, $\Delta t_\mathrm{bin}$ is the detector time bin width, and $T$ is the data collection time.

Figures \ref{fig:schematic}(c) and \ref{fig:schematic}(d) show the biphoton correlations for paths 1 and 2, respectively, measured without the presence of the two beam splitters. They are nearly identical to each other, with a coherence time of about 300 ns, corresponding to a bandwidth of 1.28 MHz. With the two beam splitters presented, the correlation  $g_{++}^{(2)}(\tau; 3\pi/2, -\pi/4)$ displays a quantum beating shown in Fig. \ref{fig:schematic}(e), as predicted in Eq. (\ref{eq:g2 function}).

\textit{Bell inequality of frequency-bin entanglement.---} As shown in Eqs. (\ref{eq:G2 function})-(\ref{eq:Coincidence}) and confirmed in our experiment, the two-photon coincidence counts are functions of the relative arrival time delay $\tau=t_\mathrm{as}-t_\mathrm{s}$ between Stokes and anti-Stokes photons at the two distant detectors and the relative phase difference $\Delta\phi_\mathrm{s}-\Delta\phi_\mathrm{as}$.  We set the photon counters D$_\mathrm{s\pm}$ enough far away from D$_\mathrm{as\pm}$, and also the two phase shifts (PS) far away, to make our Bell test locality-loophole free.  Meanwhile, we take the fair-sampling assumption. To test the Clauser-Horner-Shimony-Holt (CHSH) type Bell inequality \cite{CHSH1969}, we define the measurement output as +1 for coincidence between D$_\mathrm{s+}$ and D$_\mathrm{as+}$ (or D$_\mathrm{s-}$ and D$_\mathrm{as-}$), and -1 for for coincidence between D$_\mathrm{s+}$ and D$_\mathrm{as-}$ (or D$_\mathrm{s-}$ and D$_\mathrm{as+}$). Then the Bell correlation coefficient can be obtained from
\begin{eqnarray}
E(\tau; \Delta\phi_\mathrm{s}, \Delta\phi_\mathrm{as})=\frac{C_{++}+C_{--}-C_{+-}-C_{-+}}{C_{++}+C_{--}+C_{+-}+C_{-+}}.
\label{eq:Ecorrelation}
\end{eqnarray}
Considering the symmetries of the two SFWM paths and the beam splitters, Eq. (\ref{eq:Ecorrelation}) can be reduced to
\begin{eqnarray}
E(&\tau&; \Delta\phi_\mathrm{s}, \Delta\phi_\mathrm{as}) \nonumber\\
&=& \frac{C_{++}(\tau; \Delta\phi_\mathrm{s}, \Delta\phi_\mathrm{as})-C_{++}(\tau; \Delta\phi_\mathrm{s}^\bot, \Delta\phi_\mathrm{as})}{C_{++}(\tau; \Delta\phi_\mathrm{s}, \Delta\phi_\mathrm{as})+C_{++}(\tau; \Delta\phi_\mathrm{s}^\bot, \Delta\phi_\mathrm{as})},
\label{eq:E value}
\end{eqnarray}
which requires only two photon detectors with $\Delta\phi_s^\bot=\Delta\phi_s+\pi$. Then the CHSH Bell parameter $S$ can be estimated as
\begin{eqnarray}
S(\tau) &=& E(\tau; \Delta\phi_\mathrm{s}, \Delta\phi_\mathrm{as})-E(\tau; \Delta\phi_\mathrm{s}', \Delta\phi_\mathrm{as}) \nonumber \\
&+&E(\tau; \Delta\phi_\mathrm{s}, \Delta\phi_\mathrm{as}')+E(\tau; \Delta\phi_\mathrm{s}', \Delta\phi_\mathrm{as}').
\label{eq:S value}
\end{eqnarray}
Under local realism the Bell inequality holds $|S(\tau)|\leq2$.

For the biphoton source described by Eqs. (\ref{eq:Entangled state})-(\ref{eq:Coincidence}), setting $\Delta\phi_\mathrm{s}' = \Delta\phi_\mathrm{s}-\pi/2$ and $\Delta\phi_\mathrm{as}' = \Delta\phi_\mathrm{as}-\pi/2$, we derive \cite{SupplementalMaterial}
\begin{eqnarray}
S(\tau)=2\sqrt{2}V(\tau)\cos(\delta\tau+\Delta\phi_\mathrm{s}-\Delta\phi_\mathrm{as}+\frac{\pi}{4}),
\label{eq:S}
\end{eqnarray}
For ideal frequency-bin entanglement without accidental coincidence, i.e., $V(\tau)=1$, our theory predicts $|S|_{max}=2\sqrt{2}$, which violates the CHSH Bell inequality. It's clear that $S(\tau)$ exhibits a sinusoidal oscillation pattern with $2\sqrt{2}V(\tau)$ as the slowly varying envelope.

The measured $S(\tau)$ and Bell correlations $E(\tau)$ are shown in Fig.~\ref{fig:Bell inequality}, with the phase setting $\Delta\phi_s = 0$, $\Delta\phi_{as} = \pi/4$, $\Delta\phi_s' = -\pi/2$, and $\Delta\phi_{as}' = -\pi/4$. At $\tau=52$ ns, we have $S=-2.52\pm0.48$, which violates the classical limit. The solid theoretical curve in Fig.~\ref{fig:Bell inequality}(a) is predicted from Eq.(\ref{eq:S}) and agree well with the experiment. The oscillation amplitude follows well the dashed envelope plotted from $\pm2\sqrt{2}V(\tau)$. By adjusting the phase setting, we can violate the Bell inequality $|S|\leq 2$ for $0<\tau\leq 350$ ns, where the visibility satisfies $V>1/\sqrt{2}$.

\begin{figure}
\includegraphics[width=8cm]{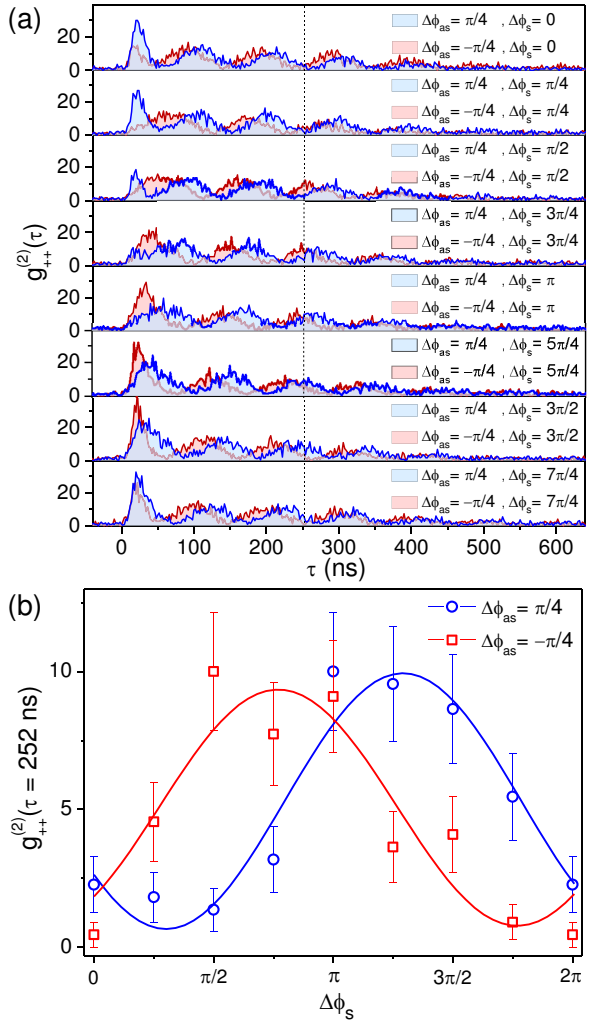}
\caption{\label{fig:NonlocalPhase} (color online). Two-photon nonlocal phase correlation. (a) The measured biphoton temporal correlations with different phase settings. (b) Phase correlations at $\tau=252$ ns. The error bars are standard deviations resulting from the statistical uncertainties of coincidence counts. The solid lines are the best fitting curves, with visibilities of $V_1 =0.86\pm0.09$ for $\Delta\phi_\mathrm{as} = \pi/4$ and $V_2 =0.85\pm0.13$ for $\Delta\phi_\mathrm{as} = -\pi/4$. }
\end{figure}

\textit{Nonlocal phase correlation.---} At a fixed relative time delay $\tau$, the two-photon correlation in Eq. (\ref{eq:G2 function}) is identical to that obtained in polarization entanglement, where $\Delta\phi_\mathrm{s}$ and $\Delta\phi_\mathrm{as}$ correspond to the orientations of distant polarizers. To confirm this, we plot  in Fig.~\ref{fig:NonlocalPhase}(a) the measured quantum beating temporal correlations with different phase settings. Figure \ref{fig:NonlocalPhase}(b) shows the nonlocal phase correlation as a function of $\Delta\phi_\mathrm{s}$ at $\tau=252$ ns under two different $\Delta\phi_\mathrm{as}$. Similarly to the polarization entanglement, the system has a rotation symmetry and the correlation depends only on the relative phase difference $\Delta\phi_\mathrm{s}-\Delta\phi_\mathrm{as}$. Therefore the visibility of the nonlocal phase correlation $V>1/\sqrt{2}$  is an indication of violation of the Bell inequality. The solid lines in Fig. \ref{fig:NonlocalPhase}(b) are the best fitting sinusoidal curves with visibilities of $V_1 =0.86\pm0.09$ for $\Delta\phi_\mathrm{as} = \pi/4$ and $V_2 =0.85\pm0.13$ for $\Delta\phi_\mathrm{as} = -\pi/4$. These visibilities are consistent with the visibility envelope of the quantum beating $V(\tau=252$ ns$)=0.78$ in Fig.~\ref{fig:schematic}e. We estimate $S$ value from $S = \sqrt{2}(V_1+V_2)= 2.42\pm0.31$, which violates the Bell inequality.

\begin{figure}
\includegraphics[width=8cm]{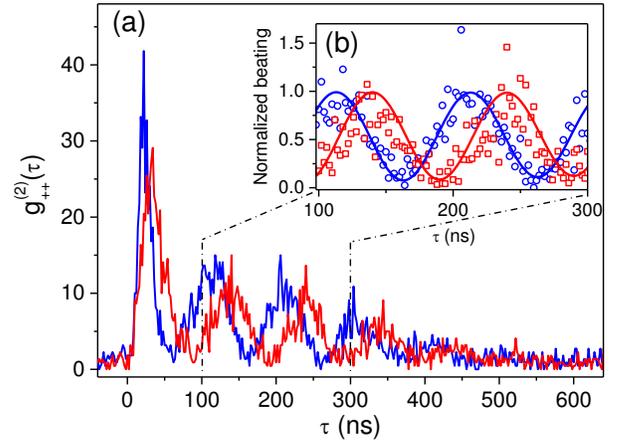}
\caption{\label{fig:quantum beating} (color online). Biphoton temporal beating.  (a) The measured two-photon temporal correlations with different phase settings. The blue data are measured with $\Delta\phi_\mathrm{s} = 3\pi/2, \Delta\phi_\mathrm{as} = -\pi/4$, and the red with $\Delta\phi_\mathrm{s} = \pi, \Delta\phi_\mathrm{as} = -\pi/4$. (b) The normalized beating signals. The solid curves are the best fittings.}
\end{figure}

\textit{Biphoton temporal beating.---} The Bell inequality was derived for general local experimental apparatus settings. For the polarization entanglement, these settings are the orientations of the polarizers. For the time-frequency entanglement, one can choose the detection time as the local detector setting parameter. In this experiment, we take $t_\mathrm{s}$ and $t_\mathrm{as}$ as the two distant local parameters. For the frequency-bin entangled state we prepared here, quantum mechanics predicts the two-photon temporal correlation exhibits a temporal quantum beating [Eq. (\ref{eq:G2 function})], which in mathematical form is similar to the polarization correlation from polarization entanglement. Figure \ref{fig:quantum beating}(a) shows the temporal beatings of two-photon correlation at two different phase settings. To compare with that in conventional polarization entanglement measurement, we normalize the quantum beating $g_{++}^{(2)}(\tau)$ to the correlation envelope $g_0^{(2)}(\tau)$ without subtracting the contribution from accidental photon coincidence counts. The normalized beating signals are plotted in Fig. \ref{fig:quantum beating}(b). With sinusoidal curve fitting for the normalized beating over 200 ns, we obtain the visibilities to be $\bar{V}=0.77\pm0.04$ and  $\bar{V}=0.76\pm0.06$, which clearly surpasses the 0.5 limit of a classical probability theory \cite{ClassicalWaveLimit1, ClassicalWaveLimit2}. The corresponding $|S|$ values are $2.17\pm0.11$ and $2.16\pm0.17$, which violate Bell inequality. 

\textit{Summary and discussion.---}In summary, we generate frequency-bin entangled narrowband biphotons from SFWM in cold atoms with a double-path configuration, where the phase difference between the two spatial paths can be controlled independently and nonlocally. The two-photon correlation exhibits a temporal quantum beating between the entangled frequency modes whose phase is determined by the relative phase difference between the two paths. We have successfully tested the CHSH Bell inequality and our best result is $|S|=2.52\pm0.48$ that violates the Bell inequality $|S|\leq2$. With $V$ as the visibility of the two-photon temporal beating, the quantum theory predicts $|S|_\mathrm{max}=2\sqrt{2}V$, which is confirmed by the experiment. Therefore the visibility $V>1/\sqrt{2}=71\%$ of the two-photon temporal beating is sufficient to violate Bell inequality. The experimental value of $|S|$ is below $2\sqrt{2}$ because the uncorrelated accidental coincidence counts from stray light, dark counts, and the randomness of photon pair generation in the spontaneous process reduces the beating visibility. We can reduce the pump laser power for lower accidental coincidence counts and thus for a higher visibility, but it will take a longer time for collecting data. The error bar of $|S|$ is the standard deviation resulting from statistical uncertainties of coincidence counts which can also be reduced with longer data taking time. The time-resolved single-photon detection is a powerful tool for quantum-state control, such as entanglement swapping and teleportation \cite{GisinNP2007}. Our result, for the first time, tests Bell inequality in nonlocal temporal correlation of frequency-bin entangled narrowband biphotons with time-resolved detection, and will have applications in quantum information processing involving time-frequency entanglement.

The work was supported by Hong Kong Research Grants Council (Project No. 16305615).



\begin{thebibliography}{99}

\bibitem{Entanglement} R. Horodecki, P. Horodecki, M. Horodecki and K. Horodecki, Quantum entanglement, Rev. Mod. Phys. \textbf{81}, 865 (2009).

\bibitem{OuPRL1988} Z. Y. Ou and L. Mandel, Violation of Bell's Inequality and Classical Probability in a Two-Photon Correlation Experiment, Phys. Rev. Lett. \textbf{61}, 50 (1988).

\bibitem{ShihPRL1988} Y. H. Shih and C. O. Alley, New type of Einstein-Podolsky-Rosen-Bohm Experiment Using Pairs of Light Quanta Produced by Optical Parametric Down Conversion, Phys. Rev. Lett. \textbf{61}, 2921 (1988).

\bibitem{KwiatPRL1995} P. G. Kwiat, K. Mattle, H. Weinfurter, A. Zeilinger, A. V. Sergienko and Y. Shih, New High-Intensity Source of Polarization-Entangled Photon Pairs, Phys. Rev. Lett. \textbf{75}, 4337 (1995).

\bibitem{HornePRL1989} M. A. Horne, A. Shimony, and A. Zeilinger, Two-Particle Interferometry, Phys. Rev. Lett. \textbf{62}, 2209 (1989).

\bibitem{RarityPRL1990} J. G. Rarity and P. R. Tapster, Experimental Violation of Bell's Inequality Based on Phase and Momentum, Phys. Rev. Lett. \textbf{64}, 2495 (1990).

\bibitem{MairNature2001} A. Mair, A. Vaziri, G. Weihs, and A. Zeilinger, Entanglement of the orbital angular momentum states of photons, Nature \textbf{412}, 313 (2001).

\bibitem{FransonPRL1989} J. D. Franson, Bell Inequality for Position and Time, Phys. Rev. Lett. \textbf{62}, 2205 (1989).

\bibitem{StefanovPRA2013} C. Bernhard, B. Bessire, T. Feurer, and A. Stefanov, Shaping frequency-entangled qudits, Phys. Rev. A \textbf{88}, 032322 (2013).

\bibitem{OlislagerPRA2010} L. Olislager, J. Cussey, A. T. Nguyen, P. Emplit, S. Massar, J.-M. Merolla, and K. P. Huy, Frequency-bin entangled photons. Phys. Rev. A \textbf{82}, 013804 (2010).

\bibitem{XieNature2015} Z. Xie, \textit{et al.} Harnessing high-dimensional hyperentanglement through a biphoton frequency comb, Nat. Photon. \textbf{9}, 536 (2015).

\bibitem{RamelowPRL2009} S. Ramelow, L. Ratschbacher, A. Fedrizzi, N. K. Langford, and A. Zeilinger, Discrete Tunable Color Entanglement, Phys. Rev. Lett. \textbf{103}, 253601 (2009).

\bibitem{GisinNP2007} M. Halder, A. Beveratos, N. Gisin, V. Scarani, C. Simon, and H. Zbinden, Entangling independent photons by time measurement, Nat. Phys. \textbf{3}, 692 (2007).

\bibitem{Bell1964} J. S. Bell, On the Einstein Podolsky Rosen paradox, Physics \textbf{1}, 195 (1964).

\bibitem{BellBook} J. S. Bell, Speakable and Unspeakable in Quantum Mechanics, (Cambridge University Press, 2004).

\bibitem{CHSH1969} J. F. Clauser, M. A. Horne, A. Shimony, and R. A. Holt, Proposed Experiment to Test Local Hidden-Variable Theories, Phys. Rev. Lett. \textbf{23}, 880 (1969).

\bibitem{EPR} A. Einstein, B. Podolsky, and N. Rosen, Can quantum-mechanical description of physical reality be considered complete? Phys. Rev. \textbf{47}, 777 (1935).

\bibitem{LoopholeFreeTest} M. Giustina, M. A. M. Versteegh, S. Wengerowsky, J. Handsteiner, A. Hochrainer, K. Phelan, F. Steinlechner, J. Kofler, J. Larsson, C. Abelln, W. Amaya, V. Pruneri, M. W. Mitchell, J. Beyer, T. Gerrits, A. E. Lita, L. K. Shalm, S. W. Nam, T. Scheidl, R. Ursin, B. Wittmann, and A. Zeilinger, Significant-Loophole-Free Test of Bell's Theorem with Entangled Photons, Phys. Rev. Lett. \textbf{115}, 250401 (2015).

\bibitem{FeketePRL2013} J. Fekete, D. Rielander, M. Cristiani, and H. de Riedmatten, Ultranarrow-Band Photon-Pair Source Compatible with Solid State Quantum Memories and Telecommunication Networks, Phys. Rev. Lett. \textbf{110}, 220502 (2013).

\bibitem{PanPRL2008} X.-H. Bao, Y. Qian, J. Yang, H. Zhang, Z.-B. Chen, T. Yang and J.-W. Pan, Generation of Narrow-Band Polarization-Entangled Photon Pairs for Atomic Quantum Memories, Phys. Rev. Lett. \textbf{101}, 190501 (2008).

\bibitem{Du2016} C. Shu, P. Chen, T. K. A. Chow, L. Zhu, Y. Xiao, M. M. T. Loy, and S. Du, Subnatural-linewidth biphotons from a Doppler-broadened hot atomic vapor cell, Nat. Commun. \textbf{7}, 12783 (2016).

\bibitem{DuPRL2008} S. Du, P. Kolchin, C. Belthangady, G. Y. Yin, and S. E. Harris, Subnatural Linewidth Biphotons with Controllable Temporal Length, Phys. Rev. Lett. \textbf{100}, 183603 (2008).

\bibitem{ZhaoOptica2014} L. Zhao, X. Guo, C. Liu, Y. Sun, M. M. T. Loy, and S. Du, Photon pairs with coherence time exceeding 1 $\mu$s, Optica \textbf{1}, 84 (2014).

\bibitem{ChenSR2015} Z. Han, P. Qian, L. Zhou, J. F. Chen, and W. Zhang, Coherence time limit of the biphotons generated in a dense cold atom cloud, Sci. Rep. \textbf{5}, 9126 (2015).

\bibitem{KurtsieferPRL2013} B. Srivathsan, G. K. Gulati, B. Chng, G. Maslennikov, D. Matsukevich, and C. Kurtsiefer, Narrow Band Source of Transform-Limited Photon Pairs via Four-Wave Mixing in a Cold Atomic Ensemble, Phys. Rev. Lett. \textbf{111}, 123602 (2013).

\bibitem{YanPRL2014} K. Liao, H. Yan, J. He, S. Du, Z.-M. Zhang and S.-L. Zhu, Subnatural-Linewidth Polarization-Entangled Photon Pairs with Controllable Temporal Length, Phys. Rev. Lett. \textbf{112}, 243602 (2014).

\bibitem{KimPRL2014}  Y.-W. Cho, K.-K. Park, J.-C. Lee, and Y.-H. Kim, Engineering Frequency-Time Quantum Correlation of Narrow-Band Biphotons from Cold Atoms, Phys. Rev. Lett. \textbf{113}, 063602 (2014).

\bibitem{TQST2015} P. Chen, C. Shu, X. Guo, M. M. T. Loy, and S. Du, Measuring the Biphoton Temporal Wave Function with Polarization-Dependent and Time-Resolved Two-Photon Interference, Phys. Rev. Lett. \textbf{114}, 010401 (2015).

\bibitem{2DMOT} S. Zhang, J. F. Chen, C. Liu, S. Zhou, M. M. T. Loy, G. K. L. Wong, and S. Du, A dark-line two-dimensional magneto-optical trap of 85Rb atoms with high optical depth, Rev. Sci. Instrum. \textbf{83}, 073102 (2012).

\bibitem{SupplementalMaterial} See the Supplemental Material, whihch includes Ref. \cite{JOSAB_Du}, for the detailed description of the experimental setup, the theory of SFWM biphoton generation from cold atoms,  and the derivation of quantum beating and CHSH Bell parameter from frequency-bin entanglement.

\bibitem{JOSAB_Du} S. Du, J. Wen, M. H. Rubin, Narrowband biphoton generation near atomic resonance, J. Opt. Soc. Am. B \textbf{25}, C98 (2008).

\bibitem{ClassicalWaveLimit1} Z. Y. Ou and L. Mandel, Violation of Bell's Inequality and Classical Probability in a Two-Photon Correaltion Experiment, Phys. Rev. Lett. \textbf{61}, 50 (1988).

\bibitem{ClassicalWaveLimit2} J.-W. Pan, D. Bouwmeester, H. Weinfurter, and A. Zeilinger, Experimental Entanglement Swapping: Entangling Photons that Never Interacted, Phys. Rev. Lett. \textbf{80}, 3891 (1998).

\end{thebibliography}
\end{document}